\documentclass[aps,pre,
superscriptaddress,groupedaddress,showpacs,twocolumn]{revtex4}
\usepackage{graphicx,epsf}
\usepackage{xcolor}
\usepackage{psfrag}
\usepackage{psfrag}
\usepackage[english]{babel}
\begin{document}
%----------------------------------------------------------------------------
%                              TITLE
%----------------------------------------------------------------------------
\title{Universal size properties of ``star-ring'' polymer structure in disordered environment}
%----------------------------------------------------------------------------
%                              AUTHORS revtex4-style
%----------------------------------------------------------------------------
\author{K. Haydukivska}
\affiliation{Institute for Condensed
Matter Physics of the National Academy of Sciences of Ukraine,\\
79011 Lviv, Ukraine}
\author{V. Blavatska}
\email[]{E-mail:  viktoria@icmp.lviv.ua}
\affiliation{Institute for Condensed
Matter Physics of the National Academy of Sciences of Ukraine,\\
79011 Lviv, Ukraine}

%----------------------------------------------------------------------------
%                             ABSTRACT
%----------------------------------------------------------------------------
\begin{abstract}
We consider the complex polymer system, consisting of ring polymer connected to the $f_1$-branched star-like structure, in good solvent
in presence of structural inhomogeneities. We assume, that structural defects are correlated at large distances $x$ according to a power law $~x^{-a}$.
Applying the direct polymer renormalization approach, we evaluate the universal size characteristics such as the ratio
of the radii of gyration of star-ring and star topologies, and compare the effective sizes of single arms in complex structures and isolated polymers of the same total molecular weight. The non-trivial impact of disorder on these quantities is analyzed.

\end{abstract}
\date{\today}
\maketitle

\section{Introduction}

In statistical description of conformational properties of long flexible polymers in a good solvent, one finds a set of characteristics, which are universal, i.e.
independent on the details of microscopic chemical structure of macromolecules \cite{deGennes,desCloiseaux}. As typical example of such properties,
we may consider the ratio of the size measures (gyration radii) of linear and closed ring polymers of the same length $L$:
 $
 g^{{\rm ring}}_{{\rm chain}}\equiv \frac{\langle R_{g\,{\rm ring}}^2\rangle}{\langle R_{g\,{\rm chain}}^2\rangle}
 $,
 which is universal
$L$-independent quantity and in the idealized case of Gaussian polymer equals $1/2$  \cite{Zimm49}.
Similarly, one can compare the size measure of a branched star-like polymer structure, consisting of $f_1$ connected arms each of length $L$ connected at one end, and  the
linear chain of the same length $f_1L$.
In the  work by Zimm and Stockmayer~\cite{Zimm49}, an estimate for the size ratio $g^{{\rm star}}_{{\rm chain}}(f_1)$ in Gaussian case was found analytically:
\begin{equation}
g^{{\rm star}}_{{\rm chain}}=\frac{3f_1-2}{f_1^2}\,. \label{zimm}
\end{equation}
Inserting $f_1=1$ or $f_1=2$ in this relation,  one restores the trivial result $g^{{\rm chain}}_{{\rm chain}}=1$. For any $f_1\geq 3$, ratio  (\ref{zimm})
is smaller than $1$, reflecting the fact that the size of a branched polymer is always smaller than the size of a linear polymer chain of the same molecular weight.

{Let us recall, that the gyration radii of all three above mentioned polymer topologies scale with length $L$ in Gaussain case according to $\langle R_{g}^2\rangle\sim L^{2\nu_{{\rm Gauss}}}$ with $\nu_{{\rm Gauss}}=1/2$.
Introducing the concept of excluded volume, which refers to the
idea that any segment (monomer) of macromolecule is not capable of occupying the space that is already occupied by another segment, leads in a good solvent regime to dimensional dependence of scaling exponent:
$\langle R_{g}^2\rangle \sim L^{2\nu_{{\rm }}(d)}$ with $\nu(d)={3}/(d+2)$ \cite{Flory53}.
Presence of excluded volume effect leads also to $d$-dependence of the size ratios   $
 g^{{\rm ring}}_{{\rm chain}}$ \cite{Baumgaertner81,Prentis82,Prentis84}
 and $g^{{\rm star}}_{{\rm chain}}$
 \cite{Daoud82,Miyake82,Miyake82_2,Alessandrini92,Whittington86,Grest87,Batouslis22,Bishop93,Wei97}.
}

\begin{figure}[b!]
\begin{center}
\includegraphics[width=73mm]{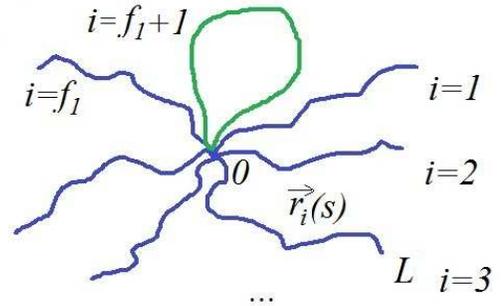}
\caption{ \label{fig:0} Schematic presentation of complex polymer structure, consisting of $f_1$ linear branches and one closed loop.}
\end{center}
\end{figure}

Both ring-like and star-like polymers play an important role both in technologies and biophysics.
In particular, one can find the circular polymers inside the living cells of bacteria \cite{Fiers62}
and higher eukaryotes \cite{Zhou03}, where DNA occurs in a closed ring shape.
Many synthetic polymers form circular structures
during polymerization and polycondensation \cite{Brown65,Geiser80,Roovers83}.
 One can encounters the star-like polymers in studying the complex systems such as gel, rubber, micellar and
other polymeric and surfactant systems~\cite{Grest96,Likos01,Ferber02}.
In the present paper, we will pay attention to the size properties of
 the ``hybrid'' complex polymer structure, consisting of $f_1$ branched linear chains connected with one closed ring (Fig. \ref{fig:0}).
 { Such a structure in particular cases $f_1$ and $f_1=2$ has a close relation with experimentally synthesized tadpole-shape polystyrene (Ref. \cite{Doi13}). Such structures are very intriguing model
polymers from the point of view of viscoelastic properties, since
an entanglement of linear parts and closed loops of different macromolecules 
could lead to formation of strong intermolecular entanglement network.
  Also, the shape properties of such tadpole structure have been analyzed numerically in \cite{Bohn10}.}
  Note that the related more general model of ``rosette-like''
polymers have been considered in the Gaussian approximation in Ref. \cite{Metzler}.
On the other hand, it is related to the process of loop formation in star polymers \cite{Haydukivska17}. It is well  known that the loop formation in macromolecules plays an important
role in a number of biochemical processes, such as stabilization of globular proteins  \cite{Perry84,Wells86,Pace88,Nagi97},
transcriptional regularization of genes \cite{Schlief88,Rippe95,Towles09},
DNA compactification in the nucleus \cite{Fraser06,Simonis06,Dorier09} etc.
Moreover, such a system can be considered as a part of a general polymer network of a more complicated structure~\cite{Duplantier89}.

In many physical processes, one faces the problem of presence of structural
obstacles (impurities) in the system.
One can encounter such situation when considering polymers in gels, colloidal solutions \cite{Pusey86},  intra-
and extracellular environments \cite{Kumarrev,cel1,cel2} etc.
 Numerous analytical and numerical studies \cite{Kremer,Grassberger93,Ordemann02,Janssen07} indicate the considerable impact of
 structural disorder on the
 effective polymer size and conformational properties of macromolecules.
 The density fluctuations of disorder often create the complex fractal structures \cite{Dullen79}.
Such situations are perfectly captured  within the frames of a model with long-range correlated
quenched defects, originally proposed in Ref. \cite{Weinrib83}.
The defects are correlated on large distances $x$
according to a power law with a pair correlation function $
h(x)\sim x^{-a}$ with $a<d$. Such a model  refers to the presence of defects of fractal structure $d_f=d-a$, with $d$ being the space dimension.
The nontrivial impact of such a type of disorder on the conformational properties of  polymers was established,
for the cases of linear chains  \cite{Blavatska01a, Blavatska10},  star-like branched polymers \cite{Blavatska06,Blavatska12} and closed ring polymers \cite{Haydukivska14}.

The present paper is dedicated to the universal size characteristics of ``star-ring'' polymer structure in solution in presence of long-range correlated disorder.
The layout of the paper is as follows. We start with presenting the continuous chain model in the next section II, then give the brief description of
the direct polymer renormalization method in Section III. The results are presented and discussed in Section IV. We end up by giving conclusions and outlook.

\section{The Model}

Within the frames of  continuous chain model \cite{Edwards}, the polymer system is considered as a set of trajectories of length $L$,  parameterized with radius vector $\vec{r}_i(s)$, $i=1,\ldots,f_1+1$
with $s$ changing from $0$ to $L$ (see Fig. \ref{fig:0}). All trajectories are considered to start at the same point, forming a structure with $f_1$ branches and one closed loop.
 The partition function of such a system  can be presented as:
\begin{eqnarray}
Z^{f_1,1}=\frac{1}{Z_0}\prod_{i=1}^{f_1+1}\!\int\!d\vec{r_i}(s)\,\delta(\vec{r}_{f_1+1}(L)-\vec{r}_{f_1+1}(0))\,{\rm e}^{-H}.
\label{ZZ}
\end{eqnarray}
{Here, $Z_0$ is partition function of Gaussian chain given by
\begin{equation}
Z_0=\prod_{i=1}^{f_1+1}\!\int\!d\vec{r_i}(s)\,{\rm e}^{-\sum\limits_{i=1}^{f_1+1}\,\int_0^L ds\,\left(\frac{d\vec{r_i}(s)}{ds}\right)^2},
\end{equation}}
$\delta$-function describes the closed loop structure and $H$ is the system Hamiltonian:
\begin{eqnarray}
&&H = \sum_{i=1}^{f_1+1}\,\int_0^L ds\,\left(\frac{d\vec{r_i}(s)}{ds}\right)^2\nonumber\\
&&+\frac{u}{2}\sum_{i,j=1}^{f_1+1}\int_0^Lds'\int_0^L ds''\,\delta(\vec{r_i}(s')-\vec{r_j}(s''))\nonumber\\
&&+\sum_{i,j=1}^{f_1+1}\int_0^Lds\,V(\vec{r_i}(s)).
\end{eqnarray}
Here, the first term describes the connectivity of trajectories, the second one corresponds to the excluded volume interactions governed by
 a coupling constant $u$ and the last term describes potential that arises due to presence of obstacles in the system.
 We consider the case when impurities are correlated on the large distances according to power law \cite{Weinrib83}:
\begin{equation}
{\overline {V(\vec{r_i}(s')) V(\vec{r_j}(s''))}} = v |\vec{r_i}(s')-\vec{r_j}(s'')|^{-a}, \label{pot}
\end{equation}
where $\overline{(\ldots)}$ denotes averaging over different realizations of disorder and $v$ is a corresponding coupling constant.

Studying the problems connected with randomness (disorder) in the system, one usually faces two types of ensemble
averaging. In so-called annealed case \cite{Brout59}, the impurity variables are a part of the disordered system phase
space, while in the quenched
case \cite{Emery75}, the free energy (the logarithm of the partition sum)
should be  averaged over an ensemble of realizations of disorder.
 In general, the critical behavior of systems with
quenched and annealed disorder is quite different.
 However, when studying the universal conformational properties of  long flexible macromolecules,
  this distinction is negligible \cite{Blavatska13} and one can use the annealed averaging, which is technically simpler.
Performing the averaging of the partition function (\ref{ZZ}) over different realizations of disorder, taking into account
up to the second moment of cumulant expansion and recalling (\ref{pot}) we obtain $\overline{ Z^{f_1,f_2} }$ in the form (\ref{ZZ})
with an effective hamiltonian:
\begin{eqnarray}
&&H_{eff} = \sum_{i=1}^{f_1+1}\,\int_0^L ds\,\left(\frac{d\vec{r_i}(s)}{ds}\right)^2\nonumber\\
&&+\frac{u}{2}\sum_{i,j=1}^{f_1+1}\int_0^Lds'\int_0^L ds''\,\delta(\vec{r_i}(s')-\vec{r_j}(s''))\nonumber\\
&&-\frac{v}{2}\sum_{i,j=1}^{f_1+1}\int_0^Lds'\int_0^L ds''\,|\vec{r_i}(s')-\vec{r_j}(s'')|^{-a}.\label{Hef}
\end{eqnarray}
Performing dimensional analysis of the  terms in (\ref{Hef}), one finds the dimensions of the couplings in terms of dimension of contour length $L$: $[u]=[L]^{(4-d)/2}$, $[v]=[L]^{(4-a)/2}$. The ``upper critical" values of the space dimension ($d_c=4$) and the correlation parameter ($a_c=4$), at which the couplings are dimensionless, play an important role in the renormalization scheme, as outlined below.

\section{The Method}
\label{Met}

The observables calculated on the basis of continuous chain model, contain divergences in the limit of infinitely long chain, that correspond to the case of infinite number of monomers. In order to receive the universal values of parameters under consideration, those divergences need to be eliminated.
The direct polymer renormalization method developed by des Cloizeaux \cite{desCloiseaux} allows to remove those divergences by adsorbing them into a
 set of so-called renormalization factors, directly connected to the observable physical quantities. The finite values of observables are obtained while evaluated at stable fixed points (FPs) of renormalization group. The method is described in more details in our previous works, e.g \cite{Blavatska12}.

It is important to note that FPs do not depend on the topology of the polymer under consideration,
and thus can be obtain in the simplest case of single linear chain.
 The renormalized coupling constants $\lambda_R=\{u_R,v_R\}$ are defined by:
\begin{eqnarray}
&&\lambda_R(\{\lambda\})= -[Z_{\{\lambda\}}(L)]^{-2}Z_{(\{ \lambda\})}(L,L)\nonumber\\
&&\times[2\pi \chi_0(L,\{\lambda_0\})]^{2-d_{\lambda}},
\end{eqnarray}
where $Z_{L}(\{\lambda_0\})$ is a partition function of a single chain, $Z_{(\{ \lambda_0\})}(L,L)$ -- partition function of two interacting chains,  $\chi_0(L,\{\lambda_0\})$ is a so-called renormalization swelling factor, and $d_{\lambda}$ are dimensions of corresponding coupling constants,
introduced after Eq. (\ref{Hef}): $d_{u}=(4-d)/2$, $d_{v}=(4-a)/2$.

In the limit of infinite linear size of macromolecules,  the renormalized theory remains finite, such that:
\begin{equation}
\lim_{L\to\infty} \lambda_{R}(\{ \lambda\})=\lambda_{R}^*\, .
\end{equation}
For negative values of ${\rm d_{\lambda}}<0$, macromolecules are expected to behave like Gaussian chains in spite of
the interactions between monomers, thus each $\lambda_{\mathrm{R}}^*=0$ for corresponding ${\rm d_{\lambda}}< 0$.
 The concept of expansion  in small
deviations from the upper critical dimensions ($\epsilon=d_c-d$, $\delta=a_c-a$) of the coupling constants thus naturally arises.
 Stable fixed points govern the asymptotical scaling properties of macromolecules in solutions and
 make it possible, e.g., to obtain the reliable
  values of universal size ratios.

\section{Results and discussions}

\subsection{Partition function}

We start our calculations by considering the partition function of the star-ring polymer structure.
We exploit the Fourier-transform of the  $\delta$-function with wave vectors $\vec{q}$ for the one corresponding to a loop structure
 and with wave vector $\vec{p_u}$ for those describing excluded volume interaction:
\begin{eqnarray}
&&\delta (\vec{r}_{f_1+1}(L)-\vec{r}_{f_1+1}(0)) =\frac{1}{(2\pi)^{d}}\times\nonumber\\
&&\times \int {\rm d}\vec{q}\, {\rm e}^{\left(-\iota\vec{q}(\vec{r}_{f_1+1}(L)-\vec{r}_{f_1+1}(0)\right)},\\ \label{d}
&&\delta (\vec{r}_i(s')-\vec{r}_j(s'')) =\frac{1}{(2\pi)^{d}} \times \nonumber\\
&&\times\int {\rm d}\vec{p_u}\, {\rm e}^{\left(-\iota\vec{p_u}(\vec{r}_i(s')-\vec{r}_j(s'')\right)}.
\end{eqnarray}
The Fourier transform of (\ref{pot}) can  be presented as:
\begin{eqnarray}
&&\frac{v}{2(2\pi)^{d}}\int\,d\vec{p_v}\,|p_v|^{d-a}{\rm e}^{-\iota\vec{p_v}(\vec{r_i}(s')-\vec{r_j}(s''))},
\end{eqnarray}
where $\iota$ is an imaginary unit. As a result, $\overline{ Z^{f_1,1}}$ can be presented as:
\begin{eqnarray}
&&\overline{ Z^{f_1,1}}=\frac{1}{Z_0}\int{\cal D}r\,\left(\frac{1}{(2\pi)^d}
{\rm e}^{-\sum\limits_{i=1}^{f_1+1}\,\int_0^L ds\,\left(\frac{d\vec{r_i}(s)}{ds}\right)^2}\right.\nonumber\\
&&\times\int\,d\vec{q}\,\,\exp(-\iota\vec{q}(\vec{r}_{f_1+1}(L)-\vec{r}_{f_1+1}(0)))\nonumber\\
&&\times\left(1-\frac{u}{2(2\pi)^{d}}\sum_{i,j=1}^{f_1+1}\int_0^Lds'\int_0^L ds''\int\,d\vec{p_u}\,\right.\nonumber\\
&&\times \exp\left(-\iota\vec{p_u}(\vec{r_i}(s')-\vec{r_j}(s''))\right)\nonumber\\
&&+\frac{v}{2(2\pi)^{d}}\sum_{i,j=1}^{f_1+1}\int_0^Lds'\int_0^L ds''\int\,d\vec{p_v}\,|p_v|^{d-a}\nonumber\\
&&\left.\left.\times\exp\left(-\iota\vec{p_v}(\vec{r_i}(s')-\vec{r_j}(s''))\right)\right)\right).\label{Zx}
\end{eqnarray}
Here, $\int{\cal D}r\ \equiv  \prod_{i=1}^{f_1+1}\,\int\,d\vec{r_i}(s)$.
Performing the corresponding integrations and  taking into account that $Z_0=(2\pi L)^{-\frac{d}{2}}$ we receive:
\begin{eqnarray}
&&\overline{ Z^{f_1,1}}=(2\pi L)^{-\frac{d}{2}}\left(1-z_u\frac{f_1\,(f_1-3)+4(f_1+1)}{\epsilon}\right.\nonumber\\
&&+z_v\frac{f_1\,(f_1-3)+4\,(f_1+1)}{\delta}-(z_u-z_v)\nonumber\\
&&\times\left.\left(2(f_1-1)-\frac{f_1(f_1-1)}{2}\ln(2)+\frac{f_1(f_1-3)}{2}\right.\right.\nonumber\\
&&\left.\left.+f_1\left(\frac{2\sqrt{5}}{5}\ln\left(\frac{2}{\sqrt{5}+3}\right)\right)\right)\right),
\end{eqnarray}
with $z_u$ and $z_v$ being dimensionless coupling constants:
\begin{eqnarray}
&&z_u=\frac{uL^{2-d/2}}{(2\pi)^{d/2}}\nonumber,\\
&&z_v=\frac{vL^{2-a/2}}{(2\pi)^{d/2}}.\label{CC}
\end{eqnarray}

\subsection{Gyration radius of a  star-ring structure and corresponding size ratios}

Gyration radius of a polymer structure under consideration in terms of continuous model can be presented as:
\begin{eqnarray}
&&\overline{\langle {R^2_{g}}\rangle} = \frac{1}{2L^2(f_1+1)^2}  \nonumber \\
&&\times \sum_{i,j=1}^{f_1+1}\int_0^L\int_0^L ds_1\,ds_2 \overline{\langle(\vec{r}_i(s_2)-\vec{r}_j(s_1))^2\rangle}.
\end{eqnarray}
Here and below, $\overline{\langle \ldots \rangle}$ denotes averaging with an effective Hamiltonian (\ref{Hef}) according to:
\begin{eqnarray}
\overline{\langle \ldots \rangle} = \frac{ \prod\limits_{i=1}^{f_1+1}\,\int\,d\vec{r_i}(s)\,\delta(\vec{r}_{f_1+1}(L)-\vec{r}_{f_1+1}(0))\,  {\rm e} ^{-H_{eff}}}{{ {{\overline{ Z^{f_1,1}}}}} }.
\nonumber
\end{eqnarray}

\begin{figure}[t!]
\begin{center}
\includegraphics[width=73mm]{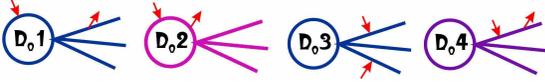}
\caption{ \label{fig:1} Diagrammatic presentation of contributions into $\langle \xi(\vec{k}) \rangle$ in Gaussian approximation. The solid line on a diagram is a schematic presentation of a polymer path of  length $L$, and
arrows denote so-called restriction points $s_1$ and $ s_2 $.}
\end{center}
\end{figure}

\begin{figure}[b!]
\begin{center}
\includegraphics[width=73mm]{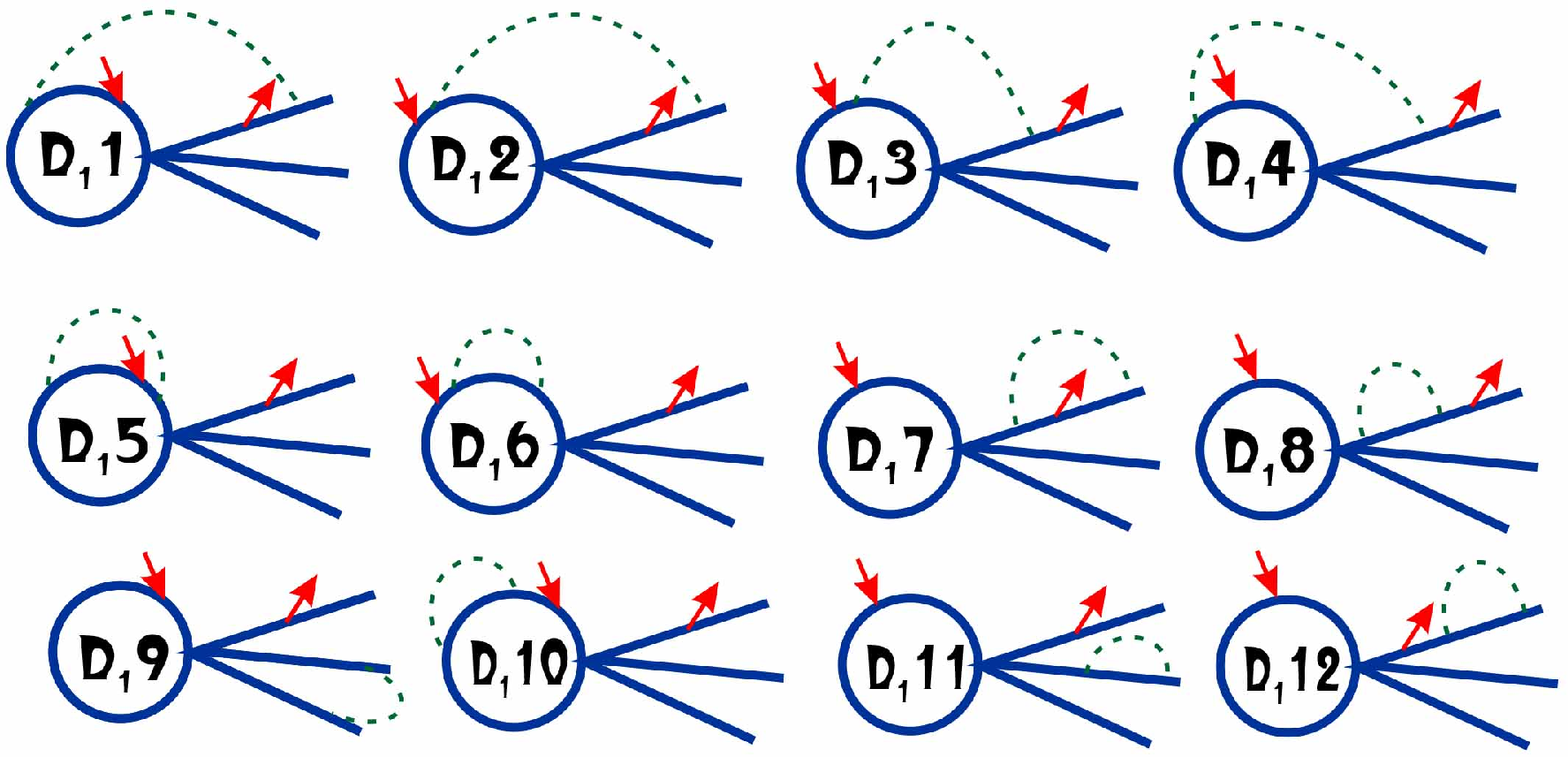}
\includegraphics[width=73mm]{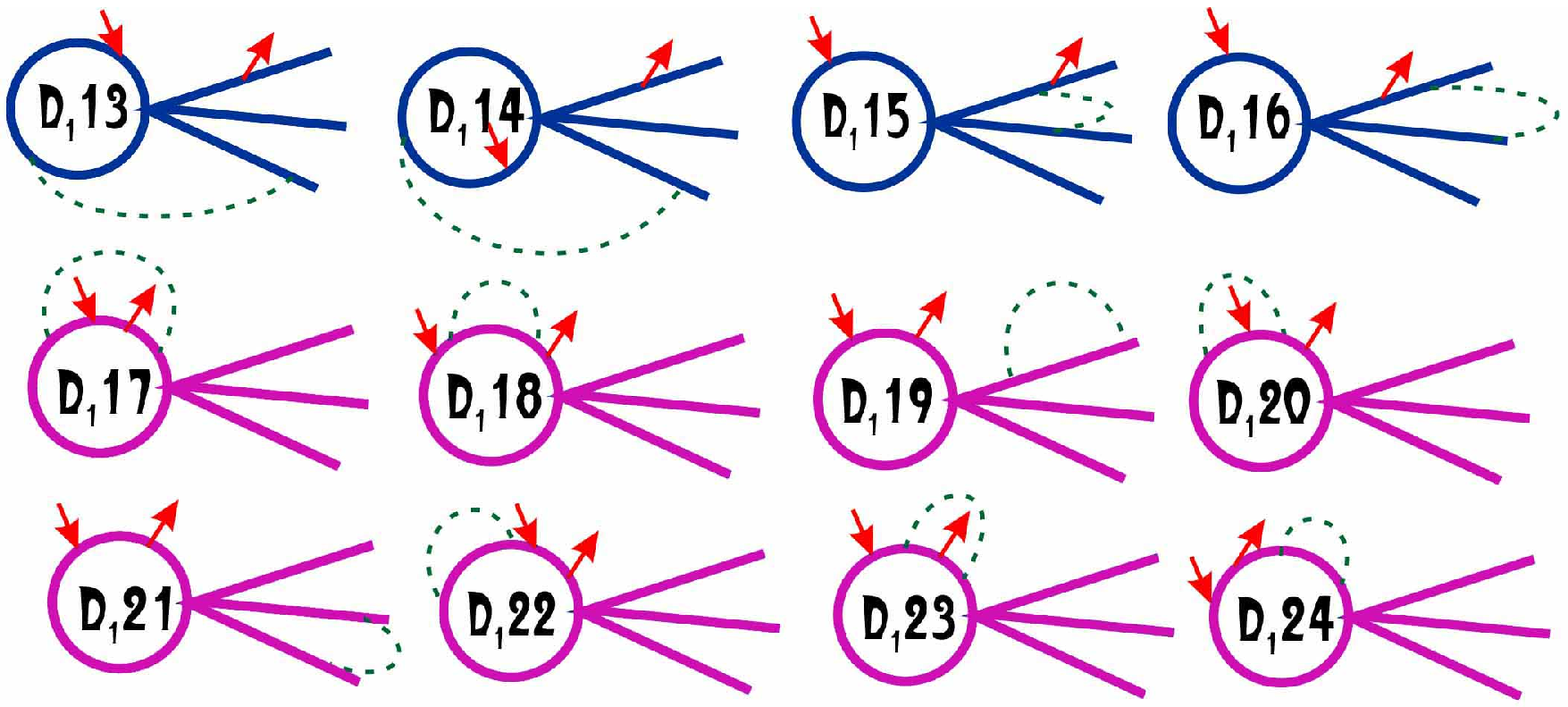}
\includegraphics[width=73mm]{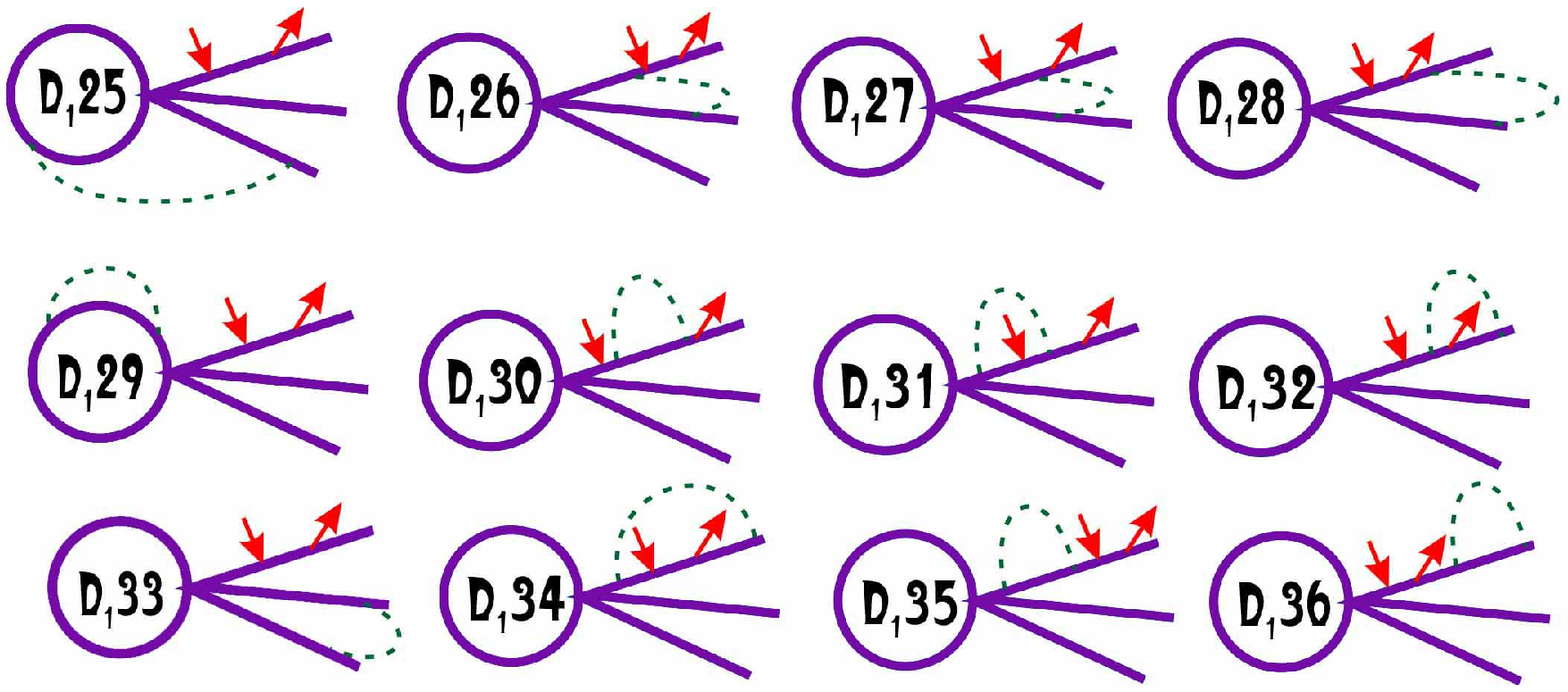}
\includegraphics[width=73mm]{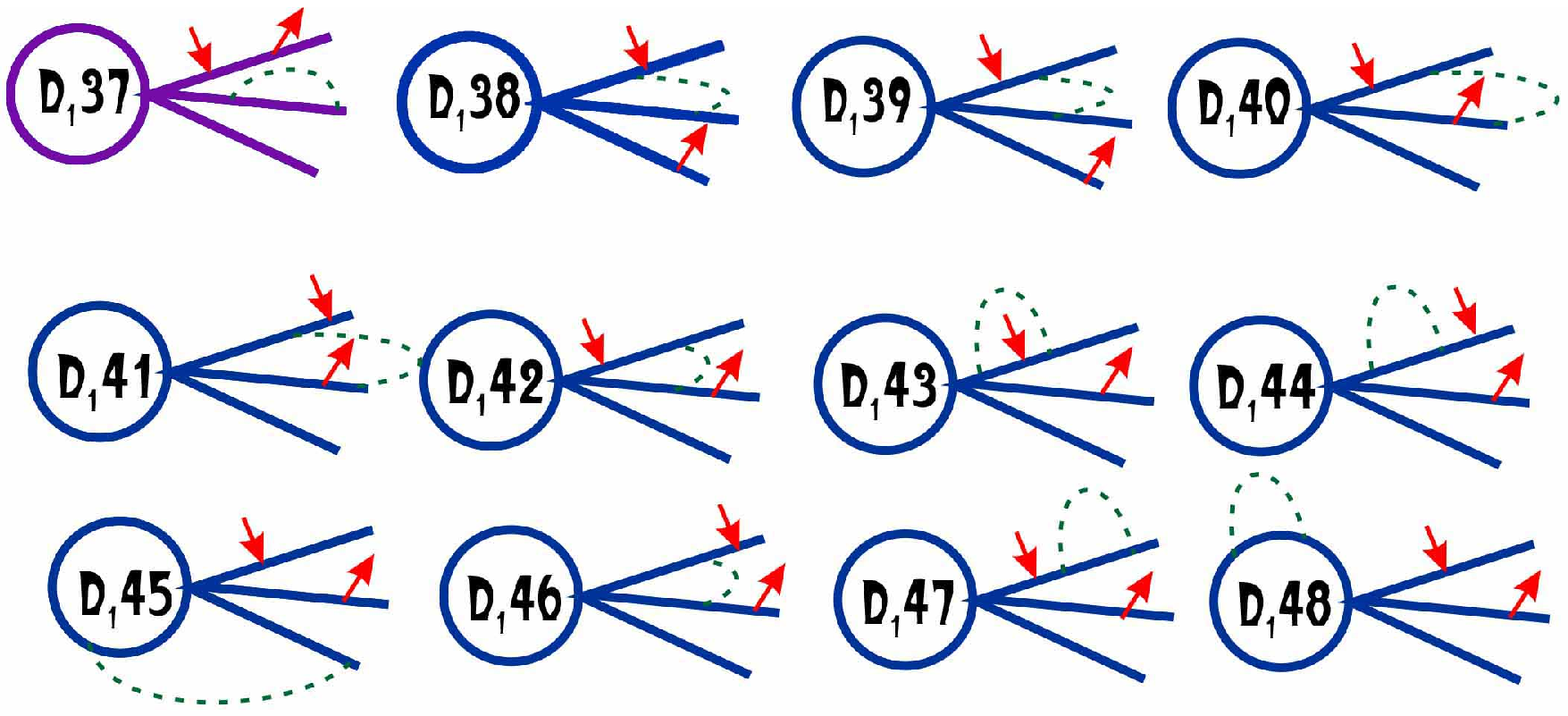}
\includegraphics[width=73mm]{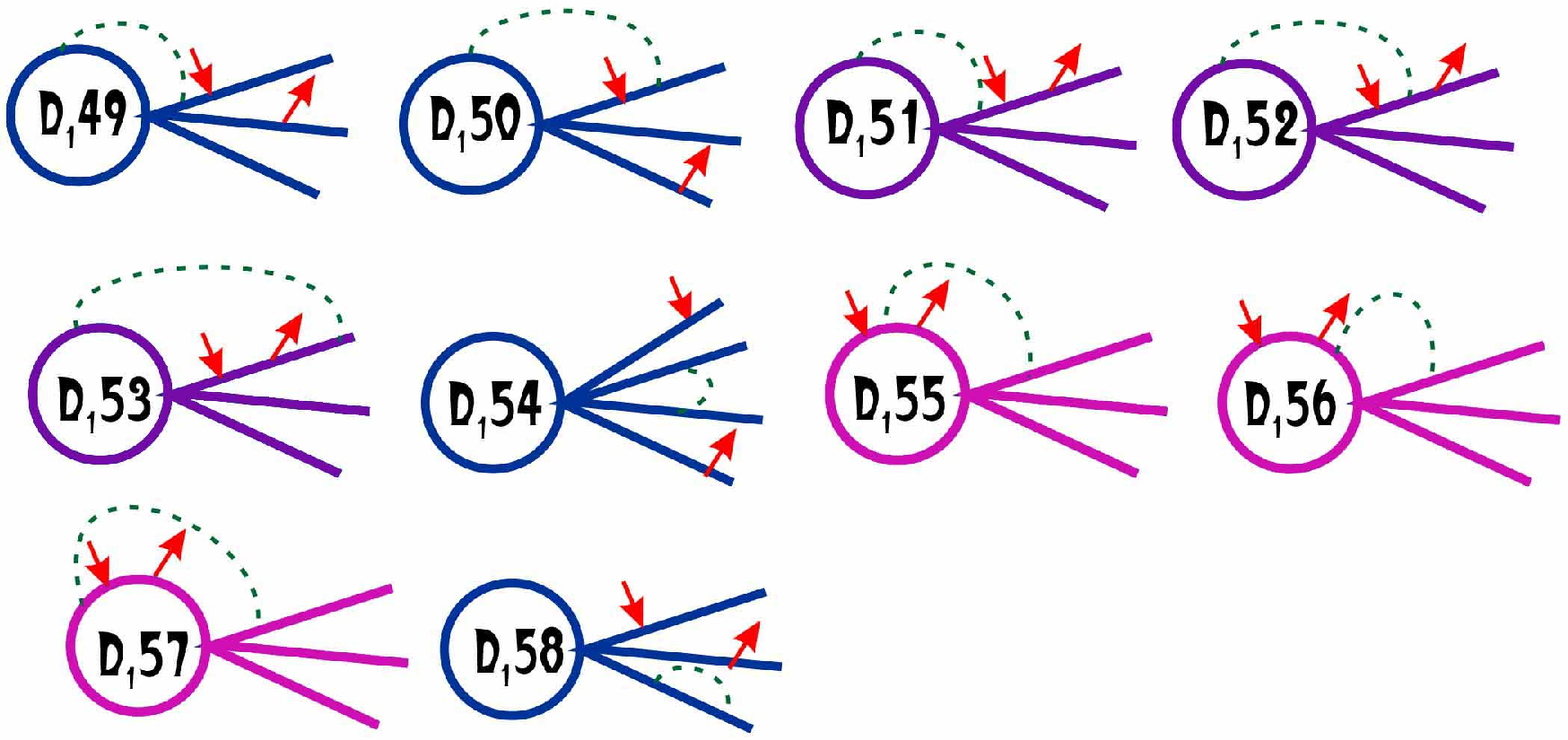}
\caption{ \label{fig:2} Diagrammatic presentation of contributions into $\langle \xi(\vec{k}) \rangle$  in the first order of perturbation theory in coupling constants. Notations are as in Fig. \ref{fig:1}; dashed lines denote the monomer-monomer interactions. Each diagram appears twice: once with excluded volume interaction governed by coupling $z_{u}$ and once with disorder interaction $z_{v}$.}
\end{center}\end{figure}

 We make use of identity:
\begin{eqnarray}
&&\overline{\langle(\vec{r}_i(s_2)-\vec{r}_j(s_1))^2\rangle} = - 2 \frac{d}{d|\vec{k}|^2}\xi(\vec{k})_{\vec{k}=0},\nonumber\\
&&\xi(\vec{k})\equiv\langle{\rm e}^{-\iota\vec{k}(\vec{r}_i(s_2)-\vec{r}_j(s_1))}\rangle, \label{der}
\end{eqnarray}
and evaluate $\xi(\vec{k})$ in path integration approach.
In calculations of the contributions into $\xi(\vec{k})$, it is convenient to use the diagrammatic presentation, as given in
 Figs. \ref{fig:1}, \ref{fig:2}.

In the simplified case of Gaussian polymer we have only four diagrams (see Fig. \ref{fig:1}). The example of diagram calculations is given in the Appendix A. It is also important to note that different diagrams are included in final expressions
 with different pre-factors arising from combinatorics, so that diagram $D_01$ is taken with pre-factor $f_1$,  diagram $D_02$ with $1$, $D_03$ with $f_1(f_1-1)/2$ and $D_04$ with $f_1$. As a result, the gyration radius of star-ring structure in Gaussian approximation reads:
\begin{equation}
\overline{\langle {R^2_{g\,{\rm Gauss}}} \rangle} = \frac{Ld (6f_1^2+6f_1+1)}{12(f_1+1)^2}.
\end{equation}

%% The general expression for a rosette-like Gaussian polymer structure with any amount of loops $f_2$ is obtained in Ref. \cite{Blavatska15}).
%%Note, that putting $f_2=0$ we may restore the gyration radius of star polymer in Gaussian approximation:
%%\begin{equation}
%%\overline{\langle {R^2_{star,{\rm Gauss} }} \rangle} = \frac{Ld(3f_1-2)}{6f_1},
%%\end{equation}
%%whereas at $f_1=0$, $f_2=1$ we obtain the expression for a single closed ring structure:
%%$\overline{\langle R^2_{ring,{\rm Gauss}} \rangle} = \frac{Ld}{12}.
%%$

In the first order of perturbation theory in couplings $z_u$, $z_v$, the gyration radius can be in general presented as:
\begin{eqnarray}
\overline{\langle R^2_{g}\rangle} = \frac{\overline{\langle R^2_{g\,{\rm Gauss}} \rangle}+z_u\overline{\langle R^2_{g\,u} \rangle}-
z_v\overline{\langle R^2_{g\,v}\rangle}}{\overline{Z^{f_1,1}}}, \nonumber
\end{eqnarray}
where $z_u,\,z_v$ are dimensionless coupling constants given by (\ref{CC}) and $\overline{\langle R^2_{g\,u} \rangle}$, $\overline{\langle R^2_{g\,v} \rangle}$ are contributions of a set of diagrams presented on Fig. \ref{fig:2} with  interactions governed by corresponding coupling constants.
Again, all the diagrams should be taken into account with corresponding combinatorial pre-factors.
Both the pre-factors and  $\epsilon,\delta$-expansions for each of the diagram are given in Table \ref{tab1} in the Appendix B.

\begin{figure}[t!]
\begin{center}
\includegraphics[width=80mm]{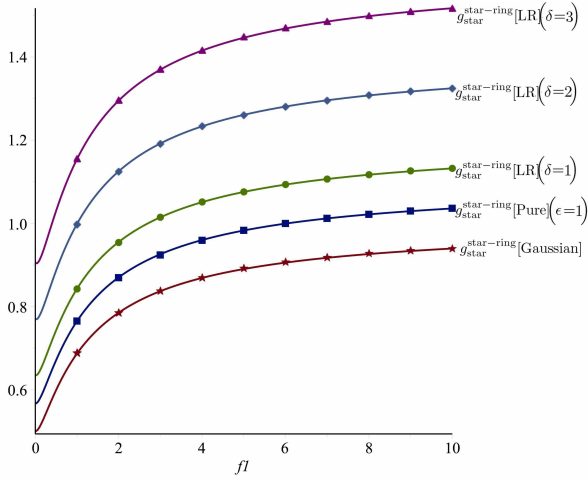}
\caption{ \label{fig:3} Size ratio $g^{{\rm star-ring}}_{{\rm star}}$ as function of parameter $f_1$, estimated at different fixed points values. Stars: Eq. (\ref{rsgaus}); squares: Eq. (\ref{rspure}) estimated at $d=3$ $(\epsilon=1)$; circles, diamonds and triangles correspond to results given by Eq. (\ref{rslr}) at different values of correlation parameter $a=3$ $(\delta=1)$, $a=2$ $(\delta=2)$, $a=1$ $(\delta=3)$ correspondingly.}
\end{center}\end{figure}

The final expression for the gyration radius of star-ring structure is thus given by:
\begin{eqnarray}
&&\overline{\langle R^2_{g}\rangle}=\frac{dL(6f_1^2+4f_1+1)}{12(f_1+1)^2}\left(1+\left(\frac{2z_u}{\epsilon}-\frac{2z_v}{\delta}\right)\right.\nonumber\\
&&-(z_u-z_v)\left(\frac{1}{30}\frac{f_1(390f_1^2-297f_1+284)}{6f_1^2+4f_1+1} \right.\nonumber\\
&&-\frac{4}{25}\frac{f_1\sqrt{5} \arctan(\sqrt{5}/5)(30f_1^2+132f_1-1)}{6f_1^2+4f_1+1}\nonumber\\
&&-\frac{8\ln(2)f_1(f_1-1)(3f_1-2)}{6f_1^2+4f_1+1}\nonumber\\
&&\left.\left.-\frac{2}{5}f_1\sqrt{5}\ln\left(\frac{2}{\sqrt{5}+3}\right)\right)\right).\label{Rf2=1}
\end{eqnarray}
Let us recall the gyration radius of a star-like polymer of the same molecular weight ($f_1+1$-arm star polymer) \cite{Blavatska12}:
\begin{eqnarray}
&&\overline{\langle R^2_{g\,{\rm star}}\rangle}=\frac{dL(3f_1+1)}{12(f_1+1)}\left(1+ \right.\nonumber  \\ &&\left.+\left(\frac{2z_u}{\epsilon}-\frac{2z_v}{\delta}\right)-\right.(z_u-z_v)\left(\frac {13}{12}\right.\nonumber\\
&&+\frac {13}{2}\frac{f_1(f_1-1)}{3f_1+1}-
\left.\left.-4\ln(2)\frac{f_1(3f_1-2)}{3f_1+1}\right)\right),\label{Rstar}
\end{eqnarray}
and the expression for the gyration radius of a single chain of the same molecular weight
(chain with total length $(f_1+1)L$) \cite{Blavatska12}:
\begin{eqnarray}
&&\langle R^2_{g\,{\rm chain}}\rangle=\frac{dL(f_1+1)}{6}\left(1+\left(\frac{2z_u}{\epsilon}-\frac{2z_v}{\delta}\right)-\right.\nonumber\\
&&\left.-(z_u-z_v)\left(\frac{13}{12} - \ln(f_1+1)\right) \right).\label{Rchain}
\end{eqnarray}

\begin{figure}[h!]
\begin{center}
\includegraphics[width=80mm]{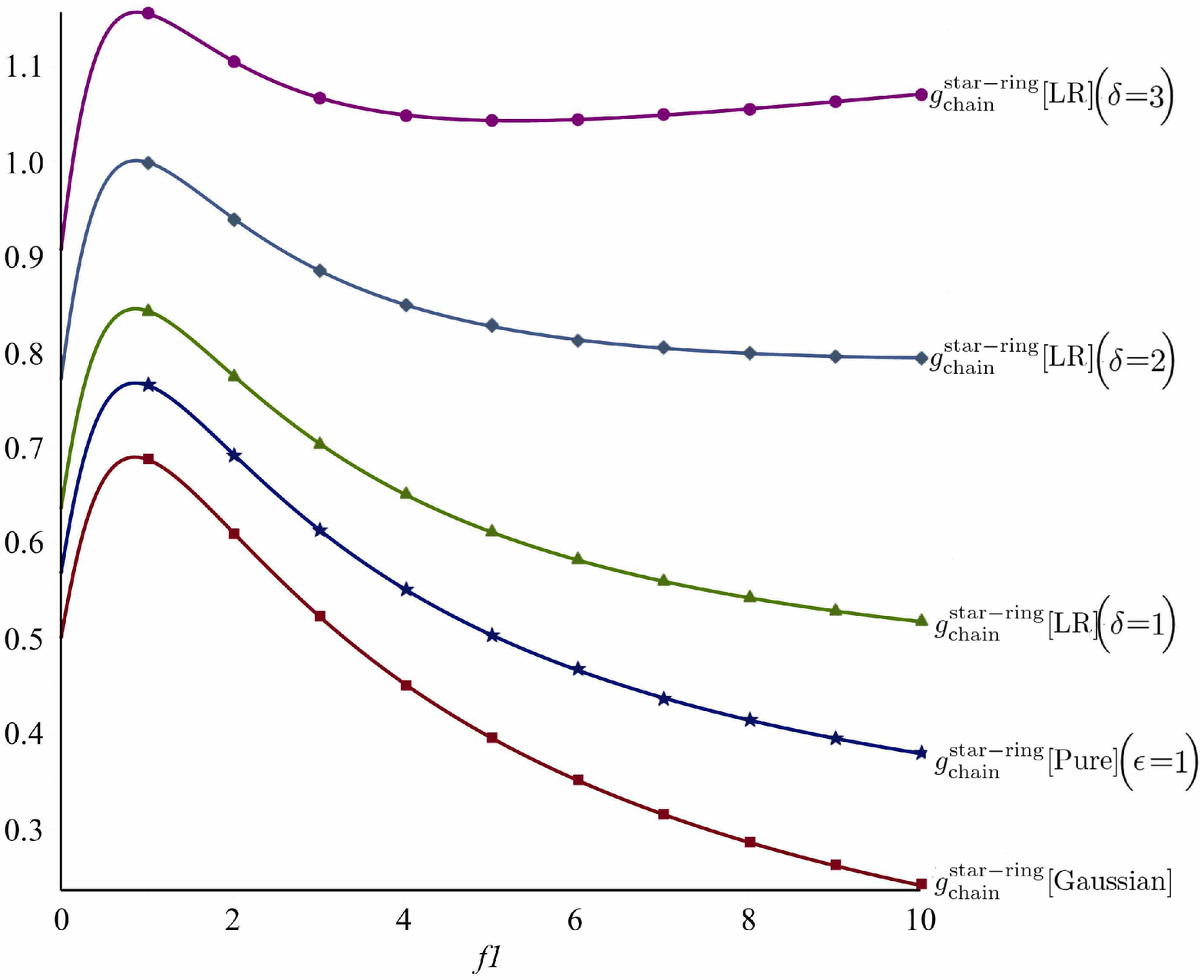}
\caption{ \label{fig:4} Size ratio $g^{{\rm star-ring}}_{{\rm chain}}$ as function of parameter $f_1$, estimated at different fixed points values. Squares: Eq. (\ref{rcgaus}); stars: Eq. (\ref{rcpure}) estimated at $d=3(\epsilon=1)$; triangles, diamonds and circles correspond to results given by Eq. (\ref{rslr}) at different values of correlation parameter $a=3$ $(\delta=1)$, $a=2$ $(\delta=2)$, $a=1$ $(\delta=3)$ correspondingly.}
\end{center}\end{figure}

With expressions  (\ref{Rf2=1})  - (\ref{Rchain}) we can obtain the corresponding universal size ratios
$g^{{\rm star-ring}}_{{\rm star}}\equiv\frac{\overline{\langle R^2_{g}\rangle}}{\overline{\langle R^2_{g\,{\rm star}}\rangle}}$ and $g^{{\rm star-ring}}_{{\rm chain}}\frac{\overline{\langle R^2_{g}\rangle}}{\overline{\langle R^2_{g\,{\rm chain}}\rangle}}$ which will allow us to estimate the relative effective size of polymers of the same molecular weight but different topology. These ratios are given by the following expressions:
\begin{eqnarray}
&&g^{{\rm star-ring}}_{{\rm star}}=\frac{6f_1^2+4f_1+1}{6f_1^2+8f_1+2}\times\nonumber\\
&&\times\left(1-(z_u-z_v)\left(\frac{4f_1(8f_1+3)(3f_1-2)\ln(2)}{(3f_1+1)(6f_1^2+4f_1+1)}\right.\right.\nonumber\\
&&-\frac{1}{60}\frac{1392f_1^3-1110f_1^2-503f_1+65}{(3f_1+1)(6f_1^2+4f_1+1)}\nonumber\\
&&-\frac{4}{25}\frac{f_1\sqrt{5}(30f_1^2+132f_1-1) \arctan(\sqrt{5}/5)}{(3f_1+1)(6f_1^2+4f_1+1)}\nonumber\\
&&\left.-\frac{2}{5}f_1\sqrt{5}\ln\left(\frac{2}{\sqrt{5}+3}\right)\right)\!,\label{g1} \\
&&\nonumber\\
&&g^{{\rm star-ring}}_{{\rm chain}}=\frac{6f_1^2+4f_1+1}{2(f_1+1)^3}\times\nonumber\\
&&\times\left(1-(z_u-z_v)\left(\frac{1}{30}\frac{f_1(390f_1^2-297f_1+284)}{6f_1^2+4f_1+1}\right.\right.\nonumber\\
&&-\frac{4}{25}\frac{f_1\sqrt{5} \arctan(\sqrt{5}/5)(30f_1^2+132f_1-1)}{6f_1^2+4f_1+1}\nonumber\\
&&-\frac{8\ln(2)f_1(f_1-1)(3f_1-2)}{6f_1^2+4f_1+1}\nonumber\\
&&\left.\left.-\frac{2}{5}f_1\sqrt{5}\ln\left(\frac{2}{\sqrt{5}+3}\right)-\frac{13}{12}+\ln(f_1+1)\right)\right)\!.\label{g2}
\end{eqnarray}

We make use of results for fixed point values found previously for the linear polymer chains
in long-range correlated disorder \cite{Blavatska01a}.  There are three distinct fixed points governing the properties of
macromolecule in various regions of parameters $d$ and $a$:
\begin{eqnarray}
&& {\rm {Gaussian}}: z^*_{u}=0,  z^*_{v}=0, \label{FPG}\\
&& { \rm {Pure}}: z^*_{u}=\frac{\epsilon}{8}, z^*_{v}=0, \label{FPP} \\
&& {\rm  {LR}}: z^*_{u}=\frac{\delta^2}{4(\epsilon-\delta)},  z^*_{v}=\frac{\delta(\epsilon-2\delta)}{4(\delta-\epsilon)}. \label{FPL}
\end{eqnarray}
Evaluating (\ref{g1}) and (\ref{g2}) in these three cases, we obtain:
\begin{eqnarray}
&&g^{{\rm star-ring}}_{{\rm star}}[{\rm Gaussian}]=\frac{6f_1^2+4f_1+1}{6f_1^2+8f_1+2}\!,\label{rsgaus}\\
&&g^{{\rm star-ring}}_{{\rm star}}[{\rm Pure}]=\frac{6f_1^2+4f_1+1}{6f_1^2+8f_1+2}\left(1-\frac{\epsilon}{8}\left(\dots\right)\right)\!,\label{rspure}\\
&&g^{{\rm star-ring}}_{{\rm star}}[{\rm LR}]=\frac{6f_1^2+4f_1+1}{6f_1^2+8f_1+2}\left(1-\frac{\delta}{4}\left(\dots\right)\right)\!,\label{rslr}\\
&&g^{{\rm star-ring}}_{{\rm chain}}[{\rm Gaussian}]=\frac{6f_1^2+4f_1+1}{2(f_1+1)^3}\!,\label{rcgaus}\\
&&g^{{\rm star-ring}}_{{\rm chain}}[{\rm Pure}]=\frac{6f_1^2+4f_1+1}{2(f_1+1)^3}\left(1-\frac{\epsilon}{8}\left(\dots\right)\right)\!,\label{rcpure}\\
&&g^{{\rm star-ring}}_{{\rm chain}}[{\rm LR}]=\frac{6f_1^2+4f_1+1}{2(f_1+1)^3}\left(1-\frac{\delta}{4}\left(\dots\right)\right)\!,\label{rclr}
\end{eqnarray}
where $\left(\dots\right)$ denotes a factor that depends only on $f_1$ and is different for different ratios.

Comparing (\ref{rsgaus}) and (\ref{rcgaus}), one easily notices, that at any $f_1$ both
 $g^{{\rm star-ring}}_{{\rm star}}[{\rm Gaussian}]$ and $g^{{\rm star-ring}}_{{\rm chain}}[{\rm Gaussian}]$ are smaller than 1.
 Thus, in Gaussian approximation the effective size of  branched polymer structure with one loop is more compact than both of that of star polymer or linear chain
 of the same molecular weight. The value of  $g^{{\rm star-ring}}_{{\rm star}}[{\rm Gaussian}]$ is growing with increasing of $f_1$ and gradually reaches the value of 1,
 which can be explained by diminishing of the role played by presence of single loop  with growing number of linear arms.
On the other hand,  $g^{{\rm star-ring}}_{{\rm chain}}[{\rm Gaussian}]$ is decreasing with $f_1$: the polymer of complex branched structure
becomes more  and more compact comparing with linear chain.
To find the quantitative values for the size ratios (\ref{rspure}), (\ref{rslr}), (\ref{rcpure}), (\ref{rclr})
we estimate them at fixed values of space dimension $d=3$ ($\epsilon=1$) and various values of correlation parameter $a$. Results are presented on Figs. \ref{fig:3}, \ref{fig:4}.
 { Note, that our results in pure solvent at $d=3$ can be compared
  with experimental values for single-tail ($f_1=1$) and
twin-tail ($f_1=2$) tadpole-shape polystyrene molecules (Ref. \cite{Doi13}): $g^{{\rm star-ring}}_{{\rm chain}}[{\rm Pure}]=0.86$ and $0.80$ correspondingly.
Note however, that our analytical results are obtained in one-loop approximation and are rather of qualitative character. To obtain the reliable estimates for the values, one need to proceed to higher order calculations and apply the special resummation techniques to the obtained perturbation theory 
expansions (see, e.g. \cite{Holovatch02}).}
Presence of excluded volume interactions as well as presence of structural disorder in the system
makes the effect of compactification of the effective size of complex branched structure less pronounced: the corresponding size ratios become closer to 1.
 It is interesting to mention, that when correlations of disorder become strong enough, the corresponding size ratios gradually overcome the
value of 1. Thus, the complex star-ring structure  becomes more extended in space, than structures without closed loops.

\subsection{Gyration radius of a single linear arm in a complex structure}

Another parameter of our interest is the gyration radius of a single linear arm within  the complex polymer structure,
which can be presented as:
\begin{eqnarray}
&&\overline{\langle {R^2_{{\rm g\, arm}}}\rangle} = \frac{1}{2L^2}\times\nonumber\\
&& \times \int_0^L\int_0^L ds_1\,ds_2 \overline{\langle(\vec{r}_1(s_2)-\vec{r}_2(s_1))^2\rangle}.
\end{eqnarray}

In this case, we need to take into account only those diagrams on Figs. \ref{fig:1} and \ref{fig:2}, which are shown in purple color, with corresponding pre-factors: $(f_1-1)$ for $D_126,\,D_127,\,D_128,\,D_137$; $(f_1-1)(f_1-2)/2$ for $D_133$; $f_2(f_1-1)$ for $D_125$ and diagrams $D_129$,  $D_151$, $D_152$, $D_153$, $D_130$, $D_131$, $D_132$, $D_134$, $D_135$, $D_136$ should be accounted for without pre-factors.
As a result we receive the following expression:
\begin{eqnarray}
&&\overline{\langle {R^2_{{\rm g\, arm}}}\rangle}=\frac{dL}{6}\left(1+\left(\frac{2z_u}{\epsilon}-\frac{2z_v}{\delta}\right)-(z_u-z_v)\right.\nonumber\\
&&\times\left(\frac{463}{120}-\frac{35}{8}f_1+{6(f_1-1)\ln(2)} \right.\nonumber\\
&&- \frac{2}{5}f_1\sqrt{5}\ln\left(\frac{2}{\sqrt{5}+3}\right)\nonumber\\
&&\left.\left.+\sqrt{5}  \arctan(\sqrt{5}/5)\frac{27-20f_1}{25}\right)\right).\label{B}
\end{eqnarray}

Let us recall the  expression for the gyration radius of a single chain of length $L$ \cite{Blavatska12}:
\begin{eqnarray}
&&\langle R^2_{{\rm g\, chain}}\rangle=\frac{dL}{6}\left(1+\left(\frac{2z_u}{\epsilon}-\frac{2z_v}{\delta}\right)-\right.\nonumber\\
&&-(z_u-z_v)\left(\frac{13}{12}\right).\label{Rc}
\end{eqnarray}

%%
%%In a case of star-like structure without closed loop we receive
%%\begin{eqnarray}
%%&&\overline{\langle {R^2_{{\rm branch}, f_1+1, 0}}\rangle}=\frac{dL}{6}\left(1+\left(\frac{2z_u}{\epsilon}-\frac{2z_v}{\delta}\right)-(z_u-z_v)\times\right.
%%\nonumber\\
%%&&\times\left.\left(\frac{131}{24}-\frac{35}{8}f_1+{\color{red}6f_1}\ln(2)\right)\right)\!\!,\label{BS}
%%\end{eqnarray}

Thus, we can consider a size ratio $g^{{\rm arm}}\equiv \overline{ \langle R^2_{{\rm g\, arm}} \rangle}/\overline{\langle R^2_{{\rm g\, chain}}\rangle}$ which reads:
\begin{eqnarray}
&&g^{{\rm arm}}=\left(1-(z_u-z_v)\right.\nonumber\\
&&\times\left(\frac{111}{40}-\frac{35}{8}f_1+{6(f_1-1)\ln(2)} - \right.\nonumber\\
&&- \frac{2}{5}f_1\sqrt{5}\ln\left(\frac{2}{\sqrt{5}+3}\right)+\nonumber\\
&&\left.\left.+\sqrt{5}  \arctan(\sqrt{5}/5)\frac{27-20f_1}{25}\right)\right). \label{arm}
\end{eqnarray}

\begin{figure}[t!]
\begin{center}
\includegraphics[width=80mm]{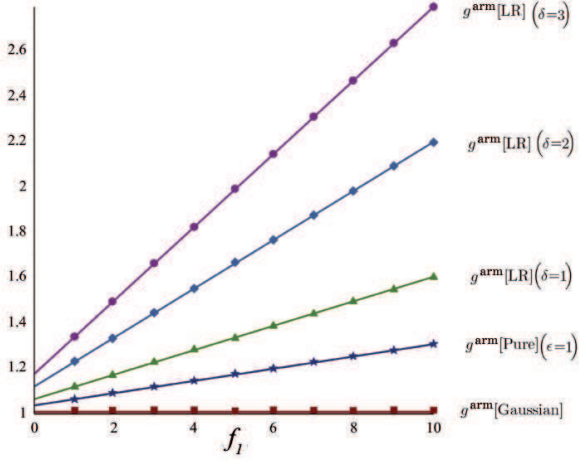}
\caption{ \label{fig:8} Size ratio $g^{{\rm arm}}$ as function of parameter $f_1$, estimated at different fixed points values. Squares: Eq. (\ref{armg}); stars: Eq. (\ref{armpure}) estimated at $d=3$ $(\epsilon=1)$; triangles, diamonds and circles correspond to results given by Eq. (\ref{armlr}) at different values of correlation parameter $a=3$ $(\delta=1)$, $a=2$ $(\delta=2)$, $a=1$ $(\delta=3)$ correspondingly.}
\end{center}\end{figure}

Substituting the fixed point values (\ref {FPG})-(\ref{FPL}) into (\ref{arm}), we obtain:
\begin{eqnarray}
&&g^{{\rm arm}}[{\rm Gaussian}]=1,\label{armg}\\
&&g^{{\rm arm}}[{\rm Pure}]=1-\frac{\epsilon}{8}\left(\dots\right),\label{armpure}\\
&&g^{{\rm arm}}[{\rm LR}]=1-\frac{\delta}{4}\left(\dots\right),\label{armlr}
\end{eqnarray}
where $\left(\dots\right)$ denotes a factor that depends only on $f_1$ and is different for different ratios.
We estimate the numerical value of (\ref{armpure}) at fixed $d=3$ ($\epsilon=1$) and
(\ref{armlr}) at
various values of correlation parameter $a$. Results are presented on Fig. \ref{fig:8}.
We note that the ratio is always  larger than 1 and grows with increasing $f_1$. Thus, the effective size of
an arm in a complex polymer structure
is more extended in space than the size of free single polymer chain.
The presence of structural disorder makes this effect more pronounced.

%{\color{red}???Numerical evaluation of coefficients in the brackets of (\ref{branch}) allow to write following this expression as:
%\begin{equation}
%g^{branch}=1-(z_u+z_v)\left(8\cdot10^{-10}f_1-0.22178\right)
%\end{equation}
%and it can be noticed that this ratio has a week, almost absent dependence on the branching parameter.
%}

\subsection{Gyration radius of a single loop in a complex structure}

\begin{figure}[b!]
\begin{center}
\includegraphics[width=80mm]{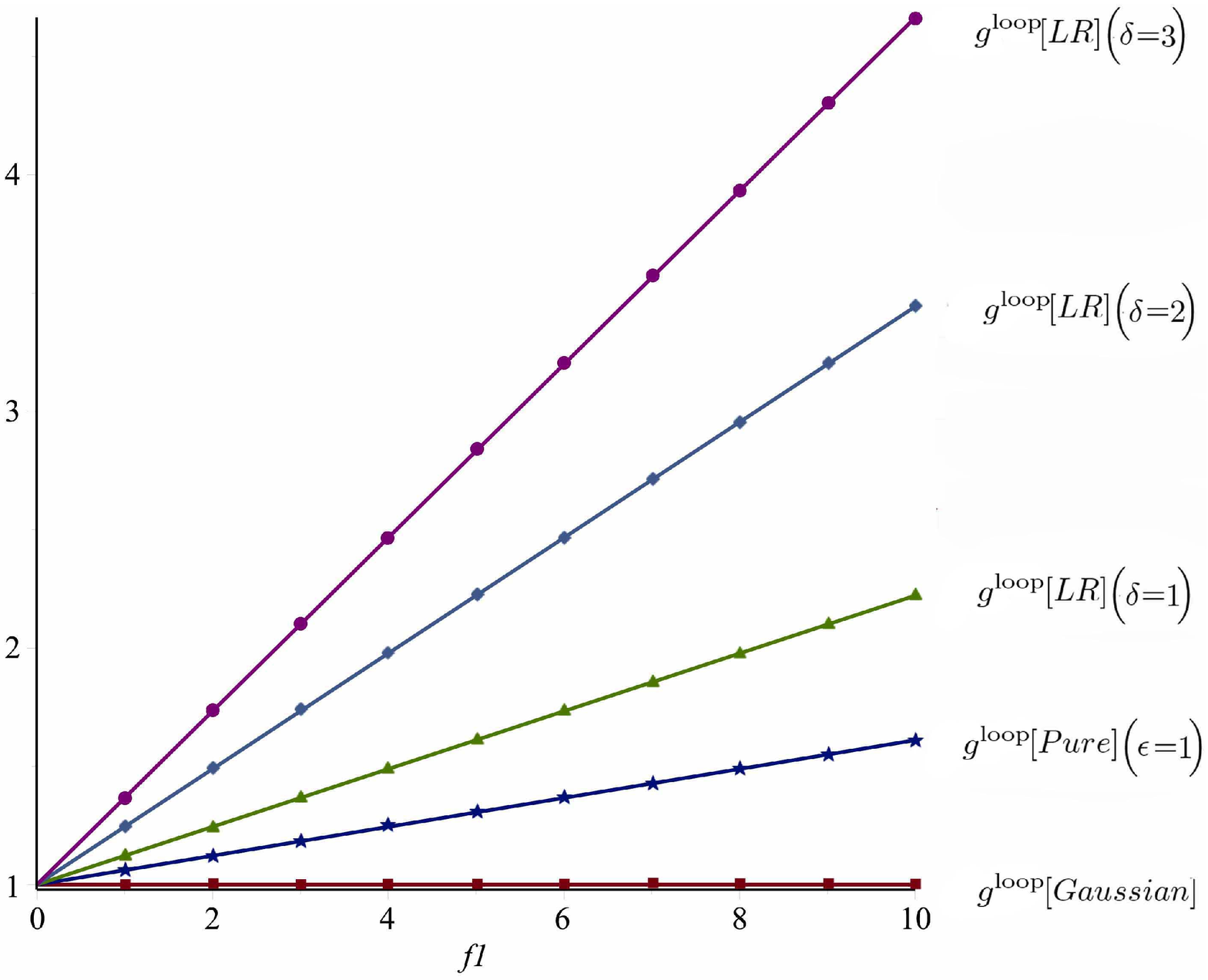}
\caption{ \label{fig:6} Size ratio $g^{{\rm loop}}$ as function of parameter $f_1$, estimated at different fixed points values. Squares: Eq. (\ref{loopgaus}); stars: Eq. (\ref{looppure}) estimated at $d=3(\epsilon=1)$; triangles, diamonds and circles correspond to results given by Eq. (\ref{looplr}) at different values of correlation parameter $a=3(\delta=1),\,a=2(\delta=2),\,a=1(\delta=3)$ corespondingly.}
\end{center}\end{figure}

In the same way as in the previous subsection,  we can calculate the gyration radius of a single loop
in a complex star-ring structure. We take into account only diagrams that marked with pink color on the figures \ref{fig:1} and \ref{fig:2}. As a result, we obtain the following expression:
\begin{eqnarray}
&&{\overline{\langle R^2_{{\rm g\, loop }}\rangle}}=\frac{dL}{12}\left(1+\left(\frac{2z_u}{\epsilon}-\frac{2z_v}{\delta}\right)-\right.\nonumber\\
&&-(z_u-z_v)\left(\frac{f_1}{5}-\frac{36\sqrt{5}f_1}{25} \arctan\left(\frac{\sqrt{5}}{5}\right)-\right. \nonumber\\
&&-\left.\frac{2 f_1\sqrt{5}}{5}\ln\left(\frac{2}{\sqrt{5}+3}\right)\right).
\label{R}
\end{eqnarray}
It is interesting to compare this result with that of gyration radius of isolated ring polymer
\cite{Haydukivska14}:
\begin{equation}
{\overline{\langle R^2_{g\,{\rm ring}}\rangle}}=\frac{dL}{12}\left(1+\left(\frac{2z_u}{\epsilon}-\frac{2z_v}{\delta}\right)\right),
\end{equation}
so that the corresponding size ratio $g^{{\rm loop}}=\frac{   {\overline{ \langle R^{2}_{g\,{\rm  loop}} \rangle}} } {  {\overline{\langle R^{2}_{g\,{\rm ring}}\rangle}} }$ reads:
\begin{eqnarray}
&&g^{{\rm loop}}=1-(z_u-z_v)\left(\frac{f_1}{5}-\right.\label{loop}\\
&&\left.-\frac{36\sqrt{5}f_1}{25} \arctan\left(\frac{\sqrt{5}}{5}\right)- \frac{2 f_1\sqrt{5}}{5}\ln\left(\frac{2}{\sqrt{5}+3}\right)\right).\nonumber
\end{eqnarray}

Evaluating (\ref{loop}) at different fixed points (\ref {FPG}) - (\ref{FPL}), we obtain:
\begin{eqnarray}
&&g^{{\rm loop}}[{\rm Gaussian}]=1,\label{loopgaus}\\
&&g^{{\rm loop}}[{\rm Pure}]=1-\frac{\epsilon}{8}(\ldots),\label{looppure}\\
&&g^{{\rm loop}}[{\rm LR}]=1-\frac{\delta}{4}(\ldots),\label{looplr}
\end{eqnarray}
where $\left(\dots\right)$ denotes a factor that depends only on $f_1$.

To find the quantitative estimates for the size ratio (\ref{looppure})
we evaluate it at fixed values of space dimension $d=3$ and various values of parameter $f_1$. Results are presented on Fig. \ref{fig:6}. We note,
that presence of excluded volume  interactions  causes the extension of effective size of a loop as a part of the complex polymer structure as comparing with isolated polymer ring.
This effect becomes more pronounced in presence of correlated disorder in system. Evaluating the size ratio (\ref{looplr}) at various fixed values of parameter $\delta$, we note an increase of this value with growing correlations of disorder.

\section{Conclusions}

In the present paper, we analyzed the universal conformational properties of complex branched polymer structure,
consisting of $f_1$ linear chains connected with one closed ring. Such a polymer system could be of interest
in processes of loop formation in branched star polymers \cite{Haydukivska17}. On the other hand,
such a system can be considered as a part of a general polymer network of a more complicated structure~\cite{Duplantier89}.

Since in most of real physical processes one encounters the problem of presence of structural
obstacles (impurities) in the system, which often have complex fractal structures \cite{Dullen79},
we turn our attention to analysis of star-ring polymer behavior in solution in presence
of defects  correlated on large distances according to a power law with a pair correlation function $
h(x)\sim x^{-a}$ with $a<d$.

Applying the direct polymer renormalization approach, we evaluate the expression for the universal ratios of the radii of gyration of star-ring structure
and star polymer (\ref{g1}) and linear chain (\ref{g2}) of the same total molecular length.
In Gaussian approximation, the effective size of  branched polymer structure with one loop is more compact than both of that of star polymer or linear chain
 of the same molecular weight.
 However, presence of excluded volume interactions as well as presence of structural disorder
makes the effect of compactification of the effective size of complex branched structure less pronounced.
Moreover, as can be seen  on Figs. \ref{fig:3}, \ref{fig:4},
when correlations of disorder become strong enough, the corresponding size ratios gradually overcome the
value of 1 and the star-ring structure becomes more extended in space, than structures without closed loops.
 Also, we analyzed the size ratio (\ref{arm}) of the radii of gyration of single arm in complex structure and free linear polymer chain of the same length.
   We found, that  the effective size of an  arm is more extended in space that of free chain,
and this extension grows in presence of long-range-correlated disorder.  Finally, we evaluate the size ratio (\ref{loop})
of the radii of gyration of a loop in complex structure and free ring polymer of the same length.
Again, we found that presence of correlated disorder in system causes the extension of effective size of a loop within the complex polymer structure as comparing with isolated polymer ring, and this effect is more pronounced with growing correlations of disorder.

\section*{Appendix A}
\begin{figure}[b!]
\begin{center}
\includegraphics[width=60mm]{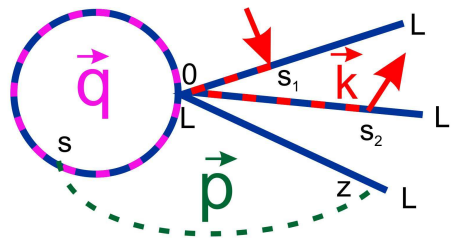}
\caption{(Color online) Example of diagrammatic contribution into the gyration radius of
star-ring polymer.}\label{fig:7}
\end{center}\end{figure}

Here, we present an example of diagram calculations. As an example we choose a diagram $D_145$ from Fig. \ref{fig:2}, which is presented in more details on Fig. \ref{fig:7}. According to the general rules of diagram calculations \cite{desCloiseaux}, each segment between any two points  and
 is oriented and bears a wave vector  given by a sum
of incoming and outcoming wave vectors injected at interaction
points and end points. Here, points $s$ and $z$ are interaction points associated with wave vector $\vec{p}$, $s_1,\,s_2$ are so-called restriction points with wave vector $\vec{k}$, the wave vector $\vec{q}$ corresponds to the loop and has restriction points at $0$ and $L$ of the corresponding trajectory. Each segment of a diagram bears a factor $-\frac{\vec{q_{ab}}^2}{2}(s_a-s_b)$ where $\vec{q_{ab}}$ is given by a sum of incoming and outcoming vectors injected at points $s_a,\,s_b$. The expression has to be integrated over all wave vectors and over all independent restriction points and end points:
\begin{eqnarray}
&&D_145 = (2\pi)^{-2d}\int^L_0\int^L_0\,ds\,dz\int^L_0\int^L_0\,ds_1\,ds_2 \times\nonumber\\
&&\times\int d\vec{q}\int d\vec{p}\,{\rm e}^{-\frac{\vec{q}^2}{2}(L-s)-\frac{(\vec{q}+\vec{p})^2}{2}s-\frac{\vec{p}^2}{2}z-\frac{\vec{k}^2}{2}(s_2+s_1)}.
\end{eqnarray}
Integrating over wave vectors $\vec{q}$ and $\vec{p}$ and taking a derivative over $\vec{k}$ according to (\ref{der}) we receive:
\begin{eqnarray}
&&D_145 = (2\pi)^{-d}L^{-d/2}\int^L_0\int^L_0\,ds_1\,ds_2 (s_2+s_1)\times\nonumber\\
&&\times\int^L_0\int^L_0\,ds\,dz \left(s+z-\frac{s^2}{L}\right)^{-d/2}.
\end{eqnarray}

\begin{table}[h!]
\begin{center}\caption{$\varepsilon$-expansion of expressions, corresponding to diagrams on Fig. \ref{fig:2}.}
\label{tab1}
  \begin{tabular}{| c | c | c | c |}
   \hline
Name  & Pre-factor &  $\epsilon$($\delta$)-expansion \\ \hline
$D_11  $ & $f_1f_2$ & $\frac{3\sqrt{5}}{100} \arctan\left( \frac{\sqrt {5}}{5} \right)+\frac{1}{10}$  \\ \hline
$D_12$ & $f_1f_2$ & $\frac{3\sqrt{5}}{100} \arctan\left( \frac{\sqrt {5}}{5} \right)+\frac{1}{10}$  \\ \hline
$D_13$ & $	f_1f_2$ & $\frac{4}{3\epsilon}-\frac{16\sqrt{5}}{15} \arctan\left( \frac{\sqrt {5}}{5}-\frac{5}{6} \right)$\\ \hline
$D_14$ & $	f_1f_2$ & $\frac{4}{3\epsilon}-\frac{16\sqrt{5}}{15} \arctan\left( \frac{\sqrt {5}}{5} \right)+\frac{5}{6}$ \\ \hline
$D_15$ & $	f_1f_2$ & $\frac{8}{3\epsilon}$ \\ \hline
$D_16$ & $	f_1f_2$ & $-\frac{1}{6\epsilon}-\frac{2}{3}$ \\ \hline
$D_17$ & $	f_1f_2$ & $\frac{4}{3\epsilon}-\frac{5}{6}$ \\ \hline
$D_18$ & $	f_1f_2$ & $-\frac{7}{3\epsilon}+\frac{1}{2}$ \\ \hline
$D_19$ & $	\frac{f_1f_2(f_1-1)(f_1-2)}{2}$ & $\frac{4}{3\epsilon}-\frac{2}{3}\ln(2)+\frac{2}{3}$ \\ \hline
$D_110$ & $	f_1f_2$ & $-\frac{1}{6\epsilon}-\frac{2}{3}$ \\ \hline
$D_111$ & $	\frac{f_1f_2(f_1-1)}{2}$ & $-\frac{4}{3\epsilon}-\frac{2}{3}$ \\ \hline
$D_112$ & $	f_1f_2$ & $-\frac{4}{3\epsilon}-\frac{1}{4}$ \\ \hline
$D_113$ & $	f_1f_2(f_1-1)$ & $\frac{4}{3\epsilon}-\frac{28\sqrt{5}}{75} \arctan\left( \frac{\sqrt {5}}{5} \right)+\frac{7}{10}$ \\ \hline
$D_114$ & $	f_1f_2(f_1-1)$ & $\frac{4}{3\epsilon}-\frac{28\sqrt{5}}{75} \arctan\left( \frac{\sqrt {5}}{5} \right)+\frac{7}{10}$ \\ \hline
$D_115$ & $	f_1f_2(f_1-1)$ & $\frac{4}{3\epsilon}+\frac{5}{3}-\frac{17}{6}\ln(2)$ \\ \hline
$D_116$ & $f_1f_2(f_1-1)$ & $\frac{1}{12}+\frac{1}{6}\ln(2)$ \\ \hline
$D_117$ & $f_2$ & $-\frac{2}{15\epsilon}-\frac{1}{30}$ \\ \hline
$D_118$ & $	f_2$ & $\frac{1}{6\epsilon}-\frac{1}{12}$ \\ \hline
$D_119$ & $	f_1f_2$ & $-\frac{1}{6\epsilon}-\frac{1}{12}$ \\ \hline
$D_120$ & $f_2$ & $\frac{1}{6\epsilon}$ \\ \hline
$D_121$ & $	\frac{f_1f_2(f_1-1)}{2}$ & $\frac{1}{6\epsilon}-\frac{1}{12}\ln(2)+\frac{1}{12}$ \\ \hline
$D_122$ & $	f_2$ & $-\frac{1}{10\epsilon}-\frac{1}{40}$ \\ \hline
$D_123$ & $	f_2$ & $\frac{1}{6\epsilon}$ \\ \hline
$D_124$ & $	f_2$ & $-\frac{1}{10\epsilon}-\frac{1}{40}$ \\ \hline
$D_125$ & $	f_1f_2(f_1-1)$ & $\frac{2}{3\epsilon}-\frac{2\sqrt{5}}{15} \arctan\left( \frac{\sqrt {5}}{5} \right)+\frac{1}{3}$ \\ \hline
$D_126$ & $	f_1(f_1-1)$ & $\frac{1}{3\epsilon}+\frac{3}{4}-\frac{3}{4}\ln(2)$ \\ \hline
$D_127$ & $	f_1(f_1-1)$ & $-\frac{4}{3}+\frac{13}{6}\ln(2)$ \\ \hline
$D_128$ & $	f_1(f_1-1)$ & $\frac{1}{48}$ \\ \hline
$D_129$ & $	f_1f_2$ & $\frac{2}{3\epsilon}-\frac{1}{3}$ \\ \hline
$D_130$ & $	f_1$ & $-\frac{2}{3\epsilon}+\frac{5}{18}$ \\ \hline
\end{tabular}
\end{center}
\end{table}

\begin{table}[b!]
\begin{center}
%\begin{table}[h!]
%\begin{center}
\caption{$\varepsilon$-expansion of expressions, corresponding to diagrams on Fig. \ref{fig:2}.}
\label{tab2}
 \begin{tabular}{| c | c | c | c |}
   \hline
%\end{tabular}
% \begin{tabular}{| c | c | c | c |}\hline

$D_131$ & $	f_1$ & $\frac{1}{3\epsilon}-\frac{5}{18}$ \\ \hline
$D_132$ & $	f_1$ & $\frac{1}{3\epsilon}-\frac{5}{18}$ \\ \hline
$D_133$ & $	\frac{f_1(f_1-1)(f_1-2)}{2}$ & $\frac{1}{3\epsilon}-\frac{1}{6}\ln(2)+\frac{1}{6}$ \\ \hline
$D_134$ & $	f_1$ & $\frac{1}{72}$ \\ \hline
$D_135$ & $	f_2$ & $-\frac{1}{3\epsilon}+\frac{5}{36}$ \\ \hline
$D_136$ & $	f_3$ & $-\frac{1}{3\epsilon}+\frac{5}{36}$ \\ \hline
$D_137$ & $	f_1(f_1-1)$ & $-\frac{1}{3\epsilon}-\frac{1}{6}$ \\ \hline
$D_138$ & $	f_1(f_1-1)(f_1-2)$ & $\frac{2}{\epsilon}+2-\frac{7}{2}\ln(2)$ \\ \hline
$D_139$ & $f_1(f_1-1)(f_1-2)$ & $\frac{1}{12}+\frac{1}{2}\ln(2)$ \\ \hline
$D_140$ & $	f_1(f_1-1)/2$ & $\frac{1}{24}$ \\ \hline
$D_141$ & $\frac{f_1(f_1-1)}{2}$ & $\frac{5}{6}-\frac{5}{6}\ln(2)$ \\ \hline
$D_142$ & $	\frac{f_1(f_1-1)}{2}$ & $\frac{5}{6}-\frac{5}{6}\ln(2)$ \\ \hline
$D_143$ & $	f_1(f_1-1)$ & $\frac{2}{\epsilon}-\frac{7}{6}$ \\ \hline
$D_144$ & $	f_1(f_1-1)$ & $-\frac{3}{\epsilon}+\frac{1}{2}$ \\ \hline
$D_145$ & $	\frac{f_1f_2(f_1-1)(f_1-2)}{2}$ & $\frac{4}{\epsilon}-\frac{4\sqrt{5}}{5} \arctan\left( \frac{\sqrt {5}}{5} \right)+2$ \\ \hline
$D_146$ & $	\frac{f_1(f_1-1)}{2}$ & $\frac{2}{\epsilon}+\frac{2}{3}-\frac{8}{3}\ln(2)$ \\ \hline
$D_147$ & $	f_1(f_1-1)$ & $-\frac{2}{\epsilon}+\frac{1}{4}$ \\ \hline
$D_148$ & $	\frac{f_1f_2(f_1-1)}{2}$ & $\frac{4}{\epsilon}-2$ \\ \hline
$D_149$ & $	f_1f_2(f_1-1)$ & $\frac{4}{\epsilon}-\frac{41\sqrt{5}}{15} \arctan\left( \frac{\sqrt {5}}{5} \right)+\frac{8}{3}$ \\ \hline  $D_150$ & $	f_1f_2(f_1-1)$ & $\frac{49\sqrt{5}}{75} \arctan\left( \frac{\sqrt {5}}{5} \right)+\frac{1}{15}$ \\ \hline
$D_151$ & $	f_1f_2$ & $\frac{2}{3\epsilon}-\frac{5\sqrt{5}}{6} \arctan\left( \frac{\sqrt {5}}{5} \right)+\frac{13}{20}$ \\ \hline
$D_152$ & $	f_1f_2$ & $\frac{13\sqrt{5}}{15} \arctan\left( \frac{\sqrt {5}}{5} \right)-\frac{3}{5}$ \\ \hline
$D_153$ & $	f_1f_2$ & $\frac{\sqrt{5}}{75} \arctan\left( \frac{\sqrt {5}}{5} \right)+\frac{1}{60}$ \\ \hline
$D_154$ & $	\frac{f_1(f_1-1)(f_1-2)(f_1-3)}{6}$ & $\frac{2}{\epsilon}-\ln(2)+1$\\ \hline
$D_155$ & $	f_1f_2$ & $\frac{\sqrt{15}}{75} \arctan\left( \frac{\sqrt {5}}{5} \right)+\frac{1}{30}$ \\ \hline
$D_156$ & $	f_1f_2$ & $\frac{1}{6\epsilon}-\frac{11\sqrt{5}}{75} \arctan\left( \frac{\sqrt {5}}{5} \right)$ \\
& & $-\frac{13}{120}$ \\ \hline
$D_157$ & $	f_1f_2$ & $\frac{1}{6\epsilon}-\frac{11\sqrt{5}}{75} \arctan\left( \frac{\sqrt {5}}{5} \right)$ \\
& & $-\frac{13}{120}$ \\ \hline
$D_158$ & $	\frac{f_1(f_1-1)(f_1-2)}{2}$ & $-\frac{2}{\epsilon}-1$ \\ \hline
\end{tabular}
\end{center}
\end{table}

Performing the integration over $s_1$ and $s_2$ and passing to dimensionless variables $\tilde{s}=s/L,\, \tilde{z}=z/L$ we come to the expression:
\begin{eqnarray}
D_145 = (2\pi)^{-d}L^{5-d}\int^1_0\int^1_0\,d\tilde{s}\,d\tilde{z} \left(\tilde{s}+\tilde{z}-\tilde{s}^2\right)^{-d/2}.
\end{eqnarray}
Integration over $\tilde{z}$ will give:
\begin{eqnarray}
&&D_145 = \frac{(2\pi)^{-d}L^{-d/2}L^{5-d/2}}{1-d/2}\times\\
&&\times\left(\int^1_0 d\tilde{s} \left(\tilde{s}+1-\tilde{s}^2\right)^{1-d/2}-\int^1_0d\tilde{s} \left(\tilde{s}-\tilde{s}^2\right)^{1-d/2}\right).\nonumber
\end{eqnarray}
Making the change of variables in first integral $\tilde{s}=t+1/2$ and performing both integrations we came to the final expression:
\begin{eqnarray}
&&D_145 = \frac{(2\pi)^{-d}L^{-d/2}L^{5-d/2}}{1-d/2}\times\nonumber\\
&&\times\left(\left(\frac{5}{4}\right)^{1-d/2}{\mbox{$_2$F$_1$}(1/2,1/2\,d-1;\,3/2;\,1/5)}\right.\nonumber\\
&&\left.-B(2-d/2,2-d/2)\right),
\end{eqnarray}
where $B$ is a Euler's Beta function and $\mbox{$_2$F$_1$}$ is a hypergeometric function.

\section*{Appendix B}

Here, we present the results of $\epsilon$-expansions for the expressions corresponding to diagrams on Fig. \ref{fig:2} (tables \ref{tab1}, \ref{tab2}). Note that in order to receive the $\delta$-expansions, one needs just to replace $\epsilon$ with $\delta$. However, it is important to note that this type of symmetry exists only in one loop approximation.


\begin{thebibliography}{99}

\bibitem{deGennes} P.G. de Gennes,  \textit{Scaling Concepts in Polymer Physics } (Cornell
University Press, Ithaca,  1979).

\bibitem {desCloiseaux} J. des Cloizeaux and G. Jannink, \textit{Polymers in Solution: Their Modelling and Structure} (Clarendon Press, Oxford, 1990).


\bibitem{Zimm49}
H. Zimm and W. H. Stockmayer, J. Chem. Phys. {\bf 17},
1301 (1949).

\bibitem {Flory53} P. Flory, \textit{Principles of Polymer Chemistry} (Cornell University Press, Ithaca, NY, 1953).


\bibitem{Baumgaertner81}
A. Baumg\"artner, J. Chem. Phys. {\bf 76}, 4275 (1982).

\bibitem{Prentis82}
J.J. Prentis, J. Chem. Phys. {\bf 76}, 1574 (1982).



\bibitem{Prentis84}
J.J. Prentis, J. Phys. A: Math. Gen. {\bf 17}, 1723 (1984).

\bibitem{Daoud82}
M. Daoud and J.P. Cotton, J. Phys. {\bf 43}, 531 (1982).

\bibitem{Miyake82}
A. Miyake and K.F.  Freed,  Macromolecules {\bf 16}, 1228 (1983).
 %
\bibitem{Miyake82_2}
A. Miyake and K.F.  Freed,  Macromolecules {\bf 17}, 678 (1984).

\bibitem{Alessandrini92}
J.L Alessandrini and M.A. Carignano, Macromolecules {\bf 25}, 1157 (1992).

\bibitem{Whittington86}
S.G. Whittington, J.E.G. Lipson,  M.K. Wilkinson and D.S. Gaunt, Macromolecules {\bf 19}, 1241 (1986).

\bibitem{Grest87}
G. Grest,  K. Kremer and T.A. Wittington, Macromolecules {\bf 20}, 1316 (1987).

\bibitem{Batouslis22}
J. Batoulis and K. Kremer, Macromolecules {\bf 22}, 4277 (1989).

\bibitem{Bishop93}
M. Bishop, J.H.R. Clarke, and J.J. Freire, J. Chem. Phys. {\bf 98}, 3452 (1993).

\bibitem{Wei97}
G. Wei, Macromolecules {\bf 30}, 2125 (1997).

\bibitem {Fiers62} W. Fiers and R.L. Sinsheimer, J.  Mol. Biol. {\bf 5},  424 (1962).

\bibitem{Zhou03} H.-X. Zhou, J. Am. Chem. Soc. {\bf 125},  9280 (2003).

\bibitem{Brown65} J.F. Brown (Jr) and G.M. Slusarczuk,  J.  Am. Chem. Soc.
{\bf 87}, 931 (1965).

\bibitem{Geiser80}
G. Geiser and H. Hocker,  Macromolecules  {\bf 13},  653 (1980).

\bibitem{Roovers83} J. Roovers and P.M. Toporowski, Macromolecules {\bf 16},  843 (1983).

\bibitem{Grest96}
 G.S. Grest, L.J. Fetters, J.S. Huang, and D. Richter,
Adv. Chem. Phys. {\bf 94}, 67 (1996).

\bibitem{Likos01} C.N. Likos, Phys. Rep. {\bf 348}, 267 (2001).

\bibitem{Ferber02} Condens. Matter Phys., 2002, \textbf{5}, No. 1. Special Issue ``Star
Polymers''. Eds. von Ferber C., Holovatch Yu.

\bibitem{Doi13}
 Y. Doi et al., Macromolecules {\bf 46}, 1075 (2013).

\bibitem{Bohn10}
M. Bohn, Heermann, O. Lourenc, and C. Cordeiro, Macromolecules {\bf 43}, 2564 (2010).

\bibitem{Metzler} V. Blavatska and R. Metzler, J. Phys. A: Math. Theor. {\bf 48}, 135001 (2015).


\bibitem{Haydukivska17} K. Haydukivska and V. Blavatska, J. Chem. Phys. {\bf 146}, 184904 (2017).



\bibitem{Perry84}
L.J. Perry and R. Wetzel, Science {\bf 226}, 555 (1984).

\bibitem{Wells86}
J.A. Wells and D.B. Powers, J. Biol. Chem. {\bf 261}, 6564 (1986).

\bibitem{Pace88}
C.N. Pace, G.R. Grimsley, J.A. Thomson, and B.J. Barnett, J. Biol. Chem. {\bf 263}, 11820 (1988).

\bibitem{Nagi97}
A.D. Nagi and L. Regan, Folding Des. {\bf 2}, 67 (1997).

\bibitem{Schlief88}
R. Schlief, Science {\bf 240}, 127 (1988).

\bibitem{Rippe95}
K. Rippe, P.H. von Hippel, and J. Langowski, Trends. Biochem. Sci. {\bf 20}, 500 (1995).

\bibitem{Towles09}
K. B. Towles, J.F. Beausang, H.G. Garcia, R. Phillips, and P.C. Nelson,  Phys. Biol. {\bf 6}, 025001 (2009).

\bibitem{Fraser06}
P. Fraser, Curr. Opin. Genet. Dev. {\bf 16}, 490 (2006)

\bibitem{Simonis06}
M. Simonis, P. Klous, E. Splinter, Y. Moshkin, R. Willemsen, E. de Wit, B. van Steensel,  and W. de Laat, Nat. Genet. {\bf 38}, 1348 (2006).

\bibitem{Dorier09}
J. Dorier and A. Stasiak, Nucl. Acids Res. {\bf 37}, 6316 (2009).



\bibitem{Duplantier89} B. Duplantier, J. Stat. Phys. {\bf 54}, 581 (1989).



\bibitem{Pusey86}P.N. Pusey and W. van Megen, Nature {\bf 320}, 340 (1986).

\bibitem{Kumarrev}S. Kumar and M.S. Li, Phys. Rep. {\bf 486}, 1 (2010).

\bibitem{cel1}F. Xiao, C. Nicholson, J. Hrabe, and S. Hrab$\breve{e}$tova, Biophys. J. {\bf 95}, 1382 (2008).

\bibitem{cel2}A.S. Verkman, Phys. Biol. {\bf 10}, 045003 (2013).



%\bibitem {Kim87}Y. Kim, J. Phys. A {\bf 20}, 1293 (1987).



\bibitem{Kremer}K. Kremer, Z. Phys. {\bf 49},  149 (1981).
\bibitem{Grassberger93}
P.~Grassberger, J. Phys. A {\bf 26}, 1023 (1993).
\bibitem{Ordemann02}
A. Ordemann, M. Porto, and H.E. Roman, Phys. Rev. E {\bf 65}, 021107 (2002).
\bibitem{Janssen07}
H.-K. Janssen  and O. Stenull, Phys. Rev. E {\bf 75}, 020801(R) (2007).

\bibitem{Dullen79}
A.L. Dullen, {\it Porous Media: Fluid Transport and Pore Structure} (Academic, New York, 1979).



\bibitem{Weinrib83}
A. Weinrib and  B.I.  Halperin,
Phys. Rev. B {\bf  27}, 413 (1983).


\bibitem{Blavatska01a}
V. Blavats'ka, C. von Ferber, and Yu. Holovatch, J. Mol. Liq. {\bf  91},
 77 (2001);
 Phys. Rev. E { \bf 64},
 041102 (2001).

\bibitem{Blavatska10}
V. Blavatska, C. von Ferber, and Yu. Holovatch,
 Phys. Lett. A {\bf 374}, 2861 (2010).


\bibitem{Blavatska06}
V. Blavatska, C. von Ferber, and Yu. Holovatch,
 Phys. Rev. E {\bf 74}, 031801  (2006).



\bibitem{Blavatska12}
V. Blavatska, C. von Ferber, and Yu. Holovatch, Condens. Matter Phys. {\bf 15}, 33603 (2012)

\bibitem{Haydukivska14} K. Haydukivska and V. Blavatska, J. Chem. Phys. {\bf 141}, 094906 (2014).

\bibitem{Edwards} S.F. Edwards, Proc. Phys. Soc. Lond. {\bf 85}, 613 (1965); Proc. Phys. Soc. Lond. {\bf 88}, 265 (1965).




\bibitem{Brout59}R. Brout,  Phys. Rev.
{\bf 115}, 824 (1959).

\bibitem{Emery75}
V. J. Emery,   Phys. Rev. B
{\bf 11},
239 (1975); S. F. Edwards  and P. W. Anderson,  J. Phys. F
{\bf 5},
965 (1975).

\bibitem{Blavatska13}
V. Blavatska, J. Phys.: Condens. Matter {\bf 25}, 505101 (2013).

\bibitem{Holovatch02}
Yu. Holovatch et al., Int. J. Mod. Phys. B {\bf 16},  4027 (2002). 

%\bibitem{Privman} V. Privman and J. Rudnick, J. Phys. A: Math. Gen. {\bf 18}, L789 (1985).



%\bibitem{Prud} A.P. Prudnikov, Yu.A. Brychkov and O.I. Marychev,
%{\em Integrals and Series: Special functions}
%(Nauka, Moskow,
%1983).
\end{thebibliography}
\end{document}